\documentclass{article}
\usepackage[T1]{fontenc} 
\usepackage[utf8]{inputenc} 
\usepackage{ismir,amsmath,cite,url}
\usepackage{booktabs} 
\usepackage{graphicx}
\usepackage{subcaption} 

\title{\vspace{9mm}Long-form music generation with latent diffusion}

\multauthor
{Zach Evans \hspace{1cm} Julian D. Parker \hspace{1cm} CJ Carr} { \bfseries{Zack Zukowski \hspace{1cm} Josiah Taylor \hspace{1cm} Jordi Pons}\vspace{5mm} \\
   Stability AI \vspace{-5mm}\\
}

\def\authorname{Z. Evans, J. D. Parker, CJ Carr, Z. Zukowski, J. Taylor and J. Pons}

\usepackage[bookmarks=false,pdfauthor={\authorname},pdfsubject={\papersubject},hidelinks]{hyperref}

\begin{document}

\maketitle

\begin{abstract}
Audio-based generative models for music have seen great strides recently, but so far have not managed to produce full-length music tracks with coherent musical structure from text prompts. We show that by training a generative model on long temporal contexts it is possible to produce long-form music of up to 4m\,45s. Our model consists of a diffusion-transformer operating on a highly downsampled continuous latent representation {(latent rate of 21.5\,Hz)}. It obtains state-of-the-art generations according to metrics on audio quality and prompt alignment, and subjective tests reveal that it produces full-length music with coherent structure.
\end{abstract}

\section{Introduction}\label{sec:introduction}

Generation of musical audio using deep learning has been a very active area of research in the last decade. 
Initially, efforts were primarily directed towards the unconditional generation of musical audio~\cite{oord2016wavenet,mehri2016samplernn}. Subsequently, attention shifted towards conditioning models directly on musical metadata~\cite{dhariwal2020jukebox}.
Recent work has focused on adding natural language control via text conditioning~\cite{agostinelli2023musiclm,schneider2023mousai, huang2023noise2music, liu2023audioldm}, and then improving these architectures in terms of computational complexity~\cite{copet2023simple, garcia2023vampnet, magnet, lam2023efficient}, quality~\cite{liu2023audioldm2, musicrl, stableaudio,pascual2023full} or controlability~\cite{stemgen, ditto, mariani2023multi, bassdiffusion}.

\begin{figure}
    \centering
    \includegraphics[scale = 0.77]{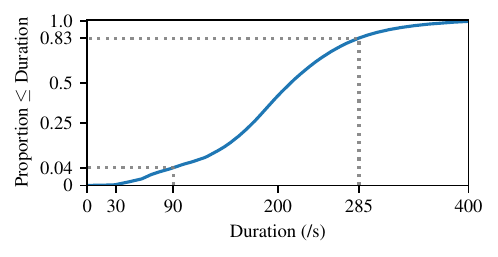}
    \vspace{-4mm}
    \caption{Cumulative histogram showing the proportion of music that is less than a particular length, for a representative sample of popular music$^1$. Dotted lines: proportion associated with the max generation length of our model~(285s) and of previous models~(90s). The vertical axis is warped with a power law for greater readability. }
    \label{fig:data}
    \vspace{-2mm}
\end{figure}

Existing text-conditioned models have generally been trained on relatively short segments of music, commonly of 10-30s in length~\cite{agostinelli2023musiclm,schneider2023mousai, huang2023noise2music, liu2023audioldm} but in some cases up to 90s~\cite{stableaudio}. These segments are usually cropped from longer compositions. Although it is possible to generate longer pieces using (e.g., autoregressive~\cite{copet2023simple}) models trained from short segments of music, the resulting music shows only local coherence and does not address long-term musical structure (see Table~\ref{tab:perceptual}, MusicGen-large-stereo results). Furthermore, the analysis of a dataset of metadata from 600k popular music tracks\footnote{{{www.kaggle.com/yamaerenay/spotify-tracks-dataset-19222021}}}(Figure~\ref{fig:data}) confirms that the majority of songs are much longer than the lengths addressed by previous works. Therefore, if we want to produce a model that can understand and produce natural musical structure, it is likely necessary to train and generate on a longer time window. We identify 285s (4m\,45s) as a target length, as it is short enough to be within reach of current deep learning architectures, can fit into the VRAM of modern GPUs, and covers a high percentage of popular music.

\begin{figure*}[ht]
    \centering
    \includegraphics[scale = 0.88]{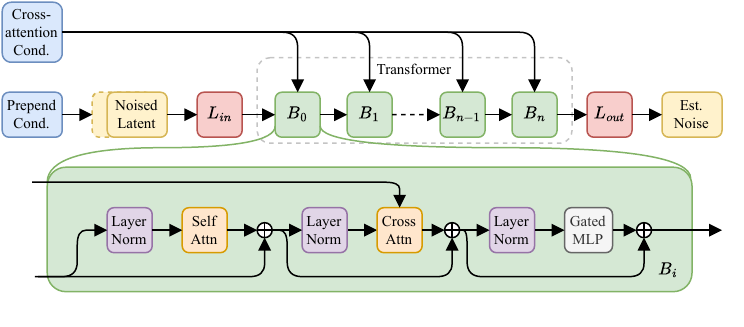}
    \vspace{-6mm}
    \caption{Architecture of the diffusion-transformer (DiT). Cross-attention includes timing and text conditioning. Prepend conditioning includes timing conditioning and also the signal conditioning on the current timestep of the diffusion process.}
    \vspace{-1mm}
    \label{fig:DiT}
\end{figure*}
\begin{figure}
    \centering
    \includegraphics[scale = 0.88]{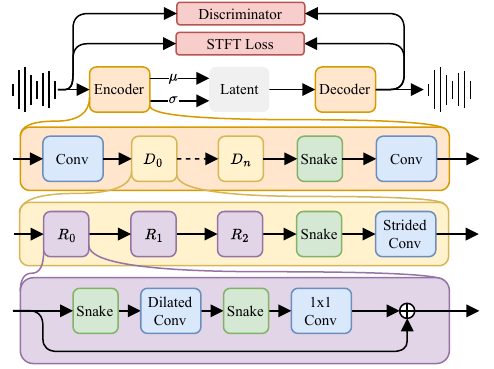}
    \vspace{-1mm}
    \caption{Architecture of the autoencoder. }
    \label{fig:VAE}
    \vspace{-3mm}
\end{figure}

 In previous works~\cite{agostinelli2023musiclm,borsos2023audiolm} it has been hypothesized that 
 ``semantic tokens enable long-term structural coherence, while modeling the acoustic tokens conditioned on the \mbox{semantic tokens enables high-quality audio synthesis''~\cite{borsos2023audiolm}.} 
 Semantic tokens are time-varying embeddings derived from text embeddings, aiming to capture the overall characteristics and evolution of music at a high level. 
 This intermediate representation is practical because it operates at low temporal resolution. 
 Semantic tokens are then employed to predict acoustic embeddings, which are later utilized for waveform reconstruction.\footnote{\cite{agostinelli2023musiclm,borsos2023audiolm} are typically conformed by four stages (denoted here as $\to$): text$\to$text-embedding$\to$semantic-token$\to$acoustic-token$\to$waveform.}Semantic tokens are commonly used in autoregressive modeling to provide guidance on what and when to stop generating~\cite{agostinelli2023musiclm,borsos2023audiolm}. 

Another line of work \cite{stableaudio} implicitly assumes that conditioning on semantic tokens is unnecessary for long-form music structure to emerge. 
Instead, it assumes that structure can emerge by training end-to-end without semantic tokens. This involves generating the entire music piece at once (full-context generation), rather than generating audio 
autoregressively guided by semantic tokens~\cite{agostinelli2023musiclm,borsos2023audiolm}. 
This approach has the potential to simplify the pipeline from four stages$^2$ to three (text$\to$text-embedding$\to$acoustic-token$\to$waveform) or even one (text$\to$waveform).
While the single-stage approach represents the closest approximation to end-to-end learning, its may be challenging to implement due to the VRAM limitations of current GPUs. Our model consists of three stages able to generate an entire music piece of 4m\,45s at once without semantic tokens.

Most music generation works rely on autoencoders to condense the long waveforms into compact latent representations (acoustic tokens or embeddings). 
Prominent examples utilize residual-vector-quantizers to provide discrete acoustic tokens~\cite{Soundstream, Encodec,DAC} for autoregressive or masked token modeling~\cite{garcia2023vampnet, stemgen, magnet, copet2023simple}.
Another prominent line of work focuses on variational autoencoders to provide a continuous and normalized acoustic embedding~\cite{schneider2023mousai,liu2023audioldm,liu2023audioldm2,stableaudio} for latent diffusion modelling. 
Our work relies on latent diffusion modeling to generate music from text prompts. Yet, and differently from prior works operating with latent rates of 40Hz to 150Hz~\cite{stableaudio,DAC,audiogen}, our autoencoder relies on a highly downsampled latent operating at 21.5Hz~(Table~\ref{tab:autoencoder}). We argue that maintaining perceptual quality at low latent rates can be essential for training generative models on long temporal contexts, enabling the creation of long-form music without the need to rely on semantic tokens. 

In our work we scale a generative model to operate over the 285s (4m\,45s) time interval. This is achieved by using a highly compressed continuous latent, and a generative model relying on latent diffusion (Sections \ref{sec:architecture} and \ref{sec:training}). 
The resulting model obtains state-of-the-art results in terms of audio quality and text-prompt coherence (Section~\ref{sec:quantitative}), and is also capable of generating long-form music with coherent structure (Sections~\ref{sec:qualitative} and \ref{sec:structure}) in 13s on a GPU.

Code to reproduce our model\footnote{{https://github.com/Stability-AI/stable-audio-tools/}} and demos \footnote{{https://stability-ai.github.io/stable-audio-2-demo/}} are online.

\section{Latent diffusion architecture}\label{sec:architecture}

Our model generates variable-length (up to 4m\,45s) stereo music at 44.1kHz from text prompts. 
It comprises three main components: an autoencoder that compresses waveforms into a manageable sequence length, a contrastive text-audio embedding model based on CLAP~\cite{elizalde2022clap, wu2023large} for text conditioning, and a transformer-based diffusion model that operates in the latent space of the autoencoder. Check their exact parametrizations online in our code repository.$^3$

\subsection{Autoencoder}

We employ an autoencoder structure that operates on raw waveforms (Figure~\ref{fig:VAE}). The encoder section processes these waveforms by a series of convolutional blocks, each of which performs downsampling and channel expansion via strided convolutions. Before each downsampling block, we employ a series of ResNet-like layers using dilated convolutions and Snake~\cite{snake} activation functions for further processing. All convolutions are parameterized in a weight-normalised form. The decoder is almost identical to the encoder structure, but employs transposed strided convolutions for upsampling and channel contraction at the start of each upsampling block. The encoder and decoder structures are similar to that of DAC~\cite{DAC}, but with the addition of a trainable $\beta$ parameter in the Snake activation, which controls the magnitude of the periodicity in the activation. We also remove the $tanh()$ activation used in DAC at the output of the decoder, as we found it introduced harmonic distortion into the signal. The bottleneck of the autoencoder is parameterized as a variational autoencoder. 

We train it using a variety of objectives. First, the reconstruction loss, consisting of a perceptually weighted multi-resolution STFT~\cite{auraloss} that deals with stereo audio as follows: the STFT loss is applied to the mid-side (M/S) representation of the stereo audio, as well as the left and right (L/R) channels separately.
The L/R component is weighted by 0.5 compared to the M/S one, and exists to mitigate potential ambiguity around L/R placement. Second, an adversarial loss term with feature matching, utilizing 5 convolutional discriminators~\cite{Encodec} with hyperparameters consistent with previous work~\cite{stableaudio}, but with channel count scaled to give $\approx$4 times  the parameter count. 
And third, the KL divergence loss term that is weighted by $\times 10^{-4}$.

\begin{table}[t]
    \centering
    \begin{tabular}{c c|c|c||c}
         & DiT & AE & CLAP  & Total \\ \midrule
          Parameters & 1.1B & 157M & 125M  & 1.3B 
    \end{tabular}
    \caption{Number of learnable parameters of our models.}
    \label{tab:params}
    \vspace{-2mm}
\end{table}

\subsection{Diffusion-transformer (DiT)}\label{sec:dit}
Instead of the widely used convolutional U-Net structure~\cite{schneider2023mousai, liu2023audioldm, liu2023audioldm2, huang2023noise2music}, we employ a diffusion-transformer~(DiT). This approach has seen notable success in other modalities~\cite{Peebles_2023_ICCV}, and has recently been applied to musical audio~\cite{apple}. The used transformer~(Figure~\ref{fig:DiT}) follows a standard structure with stacked blocks consisting of serially connected attention layers and gated multi-layer perceptrons~(MLPs), with skip connections around each. We employ layer normalization at the input to both the attention layer and the MLP. The key and query inputs to the attention layer have rotary positional embedding~\cite{rope} applied to the lower half of the embedding. Each transformer block also contains a cross-attention layer to incorporate conditioning. Linear mappings are used at the input and output of the transformer to translate from the autoencoder latent dimension to the embedding dimension of the transformer. 
We utilize efficient block-wise attention~\cite{dao2022flashattention} and gradient checkpointing~\cite{chen2016training} to reduce the computational and memory impact of applying a transformer architecture over longer sequences. These techniques are crucial to viable training of model with this context length.

The DiT is conditioned by 3 signals: \textit{text} enabling natural language control, \textit{timing} enabling variable-length generation, and \textit{timestep} signaling the current timestep of the diffusion process. Text CLAP embeddings are included via cross-attention. Timing conditioning \cite{dhariwal2020jukebox,stableaudio} is calculated using sinusoidal embeddings~\cite{ho2020denoising} and also included via cross-attention. Timing conditioning is also prepended before the transformer, along with a sinusoidal embedding describing the current timestep of the diffusion process.


\subsection{Variable-length music generation}\label{sec:variable}

Given that the nature of long-form music entails varying lengths, our model also allows for variable-length music generation. We achieve this by generating content within a specified window length (e.g., 3m\,10s or 4m\,45s) and relying on the timing condition to fill the signal up to the length specified by the user. The model is trained to fill the rest of the signal with silence. To present variable-length audio outputs shorter than the window length to end-users, one can easily trim the appended silence. We adopt this strategy, as it has shown its effectiveness in previous work \cite{stableaudio}.

\subsection{CLAP text encoder}

We rely on a contrastive model trained from text-audio pairs, following the structure of CLAP~\cite{wu2023large}. It consists of a {HTSAT-based}~\cite{chen2022hts} audio encoder with fusion and a RoBERTa-based~\cite{liu2019roberta} text encoder, both trained from scratch on our dataset with a language-audio contrastive loss. 
Following previous work~\cite{stableaudio}, we use as text features the next-to-last hidden layer of the CLAP text encoder.

\section{Training Setup}\label{sec:training}
Training the model is a multi-stage process and was conducted on a cluster of NVIDIA A100 GPUs. Firstly, the autoencoder and CLAP model are trained. The CLAP model required approximately 3k GPU hours\footnote{GPU hours represent one hour of computation on a single GPU. The training process was distributed across multiple GPUs for efficency.}and the autoencoder 16k GPU hours$^4$. Secondly, the diffusion model is trained. To reach our target length of 4m\,45s, we first pre-train the model for 70k GPU hours$^4$ on sequences corresponding to a maximum of 3m\,10s of music. We then take the resulting model and fine-tune it on sequences of up to 4m\,45s for a further 15k GPU hours$^4$. Hence, the diffusion model is first pre-trained to generate 3m\,10s music (referred to as the \textit{pre-trained} model), and then fine-tuned to generate 4m\,45s music (the \textit{fully-trained} model).

All models are trained with the AdamW optimiser, with a base learning rate of $1e-5$ and a scheduler including exponential ramp-up and decay. 
We maintain an exponential moving average of the weights for improved inference.
Weight decay, with a coefficient of 0.001, is also used.
Parameter counts for the networks are given in Table~\ref{tab:params}, and the exact hyperparameters we used are detailed online$^3$.

The DiT is trained to predict a noise increment from noised ground-truth latents, following the v-objective~\cite{salimans2022progressive}. We sample from our model using DPM-Solver++~\cite{lu2022dpm} (100 steps), with classifier-free guidance~\cite{lin2024common} (scale of 7.0).

\subsection{Training data and prompt preparation}

Our dataset consists of 806,284 files (19,500h) containing music (66\% or 94\%)\footnote{Percentages: number of files or GBs of content, respectively.}, sound effects (25\% or 5\%)$^5$, and instrument stems (9\% or 1\%)$^5$.
This audio is paired with text metadata that includes natural-language descriptions of the audio file’s contents, as well as metadata such as BPM, genre, moods, and instruments for music tracks. All of our dataset (audio and metadata) is available online\footnote{https://www.audiosparx.com/} for consultation. 
This data is used to train all three components of the system from scratch: the CLAP text encoder, the autoencoder and the DiT.
The 285s (4m\,45s) target temporal context encompasses over 90\% of the dataset.

During the training of the CLAP text encoder and the DiT, we generate text prompts from the metadata by concatenating a random subset of the metadata as a string. This allows for specific properties to be specified during inference, while not requiring these properties to be present at all times. For half of the samples, we include the metadata-type (e.g., Instruments or Moods) and join them with a delimiting character (e.g., Instruments: Guitar, Drums, Bass Guitar|Moods: Uplifting, Energetic).
For the other half, we do not include the metadata-type and join the properties with a comma (e.g., Guitar, Drums, Bass Guitar, Uplifting, Energetic). For metadata-types with a list of values, we shuffle the list. Hence, we perform a variety of random transformations of the resulting string, including two variants of delimiting character (``,'' and ``|''), shuffling orders and transforming between upper and lower case.

\section{Experiments}
\label{sec:evaluation}

\subsection{Quantitative evaluation}\label{sec:quantitative}
We evaluate a corpus of generated music using previously established metrics~\cite{stableaudio}, as implemented in \emph{stable-audio-metrics}.\footnote{https://github.com/Stability-AI/stable-audio-metrics}Those include the Fréchet distance on OpenL3 embeddings~\cite{cramer2019look}, KL-divergence on PaSST tags~\cite{koutini22passt}, and distance in LAION-CLAP space\cite{wu2023large,huang2023make}\footnote{https://github.com/LAION-AI/CLAP}.

We set MusicGen-large-stereo (MusicGen)~\cite{copet2023simple} as baseline, since it is the only publicly available model able to generate music at this length in stereo. This autoregressive model can generate long-form music of variable length due to its sequential (one-sample-at-a-time generation) sampling. However, note that MusicGen is not conditioned on semantic tokens that ensure long-term structural coherence, and it was not trained to generate such long contexts.

The prompts and ground-truth audio used for the quantitative study are from the Song Describer Dataset~\cite{manco2023song}. We select this benchmark, with 2m long music, because other benchmarks contain shorter music segments~\cite{agostinelli2023musiclm} and are inappropriate for long-form music evaluation. 
As vocal generation is not our focus and MusicGen is not trained for this task either, we opted to ensure a fair evaluation against MusicGen by curating a subset of 586 prompts that exclude vocals.\footnote{Prompts containing any of those words were removed: speech, speech synthesizer, hubbub, babble, singing, male, man, female,
woman, child, kid, synthetic singing, choir, chant, mantra, rapping, humming, groan, grunt, vocal, vocalist, singer, voice, and acapella.}This subset, referred to as the {Song Describer Dataset (no-singing)}, serves as our benchmark for comparison.
We assess 2m generations to remain consistent with the ground-truth  and also evaluate our models at their maximum generation length---which is 3m\,10s for the pre-trained model or 4m\,45s for the fully-trained one {(Tables~\ref{tab:SDDpre-train} and \ref{tab:SDDFull}, respectively)}. For each model and length we study, we generate one render per prompt in our benchmark. This results in 586 generations per experiment.

Our model is first pre-trained to generate 3m\,10s music ({pre-trained} model) and then fine-tuned to generate 4m\,45s music ({fully-trained} model). Tables~\ref{tab:SDDpre-train} and~\ref{tab:SDDFull} show the quantitative results for both models and inference times.
Comparing metrics between the pre-trained model and the fully-trained one shows no degradation, confirming the viability of extending context length via this mechanism. 
The proposed model scores better than MusicGen at all lengths while being significantly faster.


\begin{table*}[ht!]
\vspace{-9mm}
\centering
\begin{tabular}{lccccccc}
\toprule
                          &  & output  & & & & inference  \\
                          & channels/sr &  length & $\text{FD}_{openl3}$ $\downarrow$ & $\text{KL}_{passt}$ $\downarrow$ & $\text{CLAP}_{score}$ $\uparrow$ &  time \\ \midrule  
MusicGen-large-stereo \cite{copet2023simple}   & {2}/32kHz  & 2m & 204.03 & 0.49 &  0.28  &    6m\,38s \\  
{Ours (pre-trained)} & {2}/{44.1kHz} & 2m$^\dagger$  &  78.70 &  0.36 & 0.39  &    8s  \\ \midrule
MusicGen-large-stereo \cite{copet2023simple}  & {2}/32kHz  & 3m\,10s  & 213.76 & 0.50 &  0.28  &     9m\,32s  \\  
{Ours (pre-trained)} & {2}/{44.1kHz} & 3m\,10s &  89.33 &  {0.34} & 0.39  &    8s \\ 
\bottomrule
\end{tabular}
\vspace{-1mm}
\caption{\textit{Song Describer Dataset (no-singing subset):} results of the 3m\,10s pre-trained model. $^\dagger$Our pre-trained model generates 3m\,10s outputs, but during inference it can generate 2m outputs by relying on the timing conditioning. We trim audios to 2m (discarding the end silent part) for a fair quantitative evaluation against the state-of-the-art (see Section \ref{sec:variable}).}
\label{tab:SDDpre-train}
\vspace{-1mm}
\end{table*}
\begin{table*}[th]
\centering
\begin{tabular}{lccccccc}
\toprule
                          &  & output  & & & & inference  \\
                          & channels/sr &  length & $\text{FD}_{openl3}$ $\downarrow$ & $\text{KL}_{passt}$ $\downarrow$ & $\text{CLAP}_{score}$ $\uparrow$ &  time \\ \midrule  
MusicGen-large-stereo  \cite{copet2023simple} & {2}/32kHz  & 2m & 204.03 & 0.49 &  0.28  &    6m\,38s \\  
{Ours (fully-trained)} & {2}/{44.1kHz} & 2m$^\dagger$ & 79.09 &  0.35 & 0.40  &    13s \\ \midrule  
MusicGen-large-stereo \cite{copet2023simple}  & {2}/32kHz  & 4m\,45s  & 218.02 & 0.50 &  0.27  & 12m\,53s  \\  
{Ours (fully-trained)} & {2}/{44.1kHz} & 4m\,45s &  81.96 & 0.34 & 0.39  & 13s  \\ 
\bottomrule
\end{tabular}
\vspace{-1mm}
\caption{\textit{Song Describer Dataset (no-singing subset):} results of the 4m\,45s fully-trained model. $^\dagger$Our fully-trained model generates 4m\,45s outputs, but during inference it can generate 2m outputs by relying on the timing conditioning. We trim audios to 2m (discarding the end silent part) for a fair quantitative evaluation against the state-of-the-art (see Section \ref{sec:variable}).}
\label{tab:SDDFull}
\vspace{-1mm}
\end{table*}

\subsection{Qualitative evaluation} \label{sec:qualitative}
We evaluate the corpus of generated music qualitatively, with a listening test developed with webMUSHRA~\cite{schoeffler2018webmushra}. Mixed in with our generated music are generations from MusicGen and also ground-truth samples from the Song Describer Dataset (no-singing). Generations of our fully-trained model are included at both 4m\,45s and 2m long, whilst ground-truth is only available at 2m. We selected two samples from each use case that were competitive for both models. For MusicGen it was difficult to find coherently structured music, possibly because it is not trained for long-form music generation. For our model, we found some outstanding generations that we selected for the test. Test material is available on our demo page.

Test subjects were asked to rate examples on a number of qualities including audio quality, text alignment, musical structure, musicality, and stereo correctness.
We report mean opinion scores (MOS) in the following scale: \textit{bad}~(1), \textit{poor}~(2), \textit{fair}~(3), \textit{good}~(4), \textit{excellent}~(5). We observed that assessing stereo correctness posed a significant challenge for many users. To address this, we streamlined the evaluation by seeking for a binary response, correct or not, and report percentages of stereo correctness. 
All 26 test subjects used studio monitors or headphones, and self-identified as music producers or music researchers. In order to reduce test time and maximise subject engagement, we split the test into two parts. Each participant can choose one of the parts, or both, depending on their available time. 

Results in Table~\ref{tab:perceptual} indicate that the generations from our system are comparable to the ground-truth in most aspects, and superior to the existing baseline. Our model obtains \textit{good} (4) MOS scores across the board and stereo correctness scores higher than 95\%, except for 2m long generations where its musical structure is \textit{fair} (3). 
Differently from our quantitative results in Table~\ref{tab:SDDFull}, qualitative metrics show that 2m long generations are slightly worse than the 4m\,45s generations (specially musical structure). We hypothesize that this could be due to the relative scarcity of full-structured music at this length in our dataset, since most music at this length might be repetitive loops.
These results confirm that semantic tokens are not strictly essential for generating music with structure, as it can emerge through training with long contexts. Note, however, that the temporal context must be sufficiently long to obtain structured music generation. It was not until we scaled to longer temporal contexts (4m\,45s), that we observed music with \textit{good} strcuture, reflecting the inherent nature of the data.
It is also noteworthy that the perceptual evaluation of structure yields to a wide diversity of responses, as indicated by the high standard deviations in Table~\ref{tab:perceptual}. This highlights the challenge of evaluating subjective musical aspects.
Finally, MusicGen achieves a stereo correctness rate of approximately 60\%. This may be attributed to its tendency to generate mixes where instruments typically panned in the center (such as bass or kick) are instead panned to one side, creating an unnaturally wide mix that was identified as incorrect by the music producers and researchers participating in our test.


\begin{table*}[h!]
\centering
\begin{tabular}{lccc|cc}
\toprule
{Results with the fully-trained model:}& \multicolumn{3}{c|}{2m long} & \multicolumn{2}{c}{4m\,45s long} \\ \midrule
& Stable & \multicolumn{1}{|c|}{MusicGen-} & \multicolumn{1}{|c}{ground} & \multicolumn{1}{|c|}{Stable} & \multicolumn{1}{|c}{MusicGen-} \\
& Audio 2  & \multicolumn{1}{|c|}{large-stereo}& \multicolumn{1}{|c}{truth}  &  \multicolumn{1}{|c|}{Audio 2}  & \multicolumn{1}{|c}{large-stereo} \\ \midrule      
Audio Quality & 4.0$\pm$0.6 & 2.8$\pm$0.8 & 4.6$\pm$0.4 & 4.5$\pm$0.4 & 2.8$\pm$0.8 \\  
Text Alignment &  4.3$\pm$0.7 & 3.1$\pm$0.8 & 4.6$\pm$0.5 & 4.6$\pm$0.4 & 2.9$\pm$1.0 \\  
Structure  & 3.5$\pm$1.3 & 2.4$\pm$0.7 & 4.3$\pm$0.8 & 4.0$\pm$1.0 & 2.1$\pm$0.7\\  
Musicality & 4.0$\pm$0.8 & 2.7$\pm$0.9 & 4.6$\pm$0.5 & 4.3$\pm$0.7 & 2.6$\pm$0.7\\   
Stereo correctness  & 96\% & 61\% & 96\% & 100\% & 57\%\\  
\bottomrule
\end{tabular}
\vspace{-1mm}
\caption{\textit{Qualitative results}. Top: mean opinion score $\pm$ standard deviation. Bottom: percentages.}
\label{tab:perceptual}
\vspace{-1mm}
\end{table*}
\begin{table*}[th!]
\centering
\begin{tabular}{lcccccc}
\toprule
                                                          & sampling & STFT & MEL &  & latent    &   latent  \\   
                                                         &rate & distance $\downarrow$ & distance $\downarrow$ & SI-SDR $\uparrow$  &  rate&     (channels)\\ \midrule                                                            

DAC~\cite{DAC} & 44.1kHz & 0.96    & 0.52   &     10.83  &     86\,Hz &     discrete\\ 
AudioGen~\cite{audiogen} & 48kHz   &  1.17   & 0.64   &     9.27  &     50\,Hz &     discrete\\  
Encodec~\cite{Encodec,copet2023simple} & 32kHz  &  1.82   & 1.12   &    5.33  &    50\,Hz &     discrete\\ \midrule 
AudioGen~\cite{audiogen} & 48kHz &  1.10    & 0.64   &     8.82  &     100\,Hz &     continuous (32)\\  
{Stable Audio}\cite{stableaudio} & 44.1kHz &  1.19    & 0.67    &     8.62   &     43\,Hz &     continuous (64)\\ 
Ours & 44.1kHz  &  1.19    & 0.71    &     7.14  &     21.5\,Hz&     continuous (64) \\ 
\bottomrule
\end{tabular}
\vspace{-1mm}
\caption{\textit{Autoencoder reconstructions on the Song Describer Dataset (all the dataset)}. Although different autoencoders operate at various sampling rates, the evaluations are run at 44.1kHz bandwidth for a fair comparison. Sorted by latent rate.}
\label{tab:autoencoder}
\vspace{-4mm}
\end{table*}

%








\subsection{Autoencoder evaluation}
We evaluate the audio reconstruction quality of our autoencoder in isolation, as it provides a quality ceiling for our system. We achieve this by comparing ground-truth and reconstructed audio via a number of established audio quality metrics\cite{DAC,Encodec}: STFT distance, MEL distance and SI-SDR (as in AuraLoss library~\cite{auraloss}, with its default parameters). The reconstructed audio is obtained by encoding-decoding the ground-truth audio from the Song Describer Dataset (all the dataset, 706 tracks) through the autoencoder. As a comparison, we calculate the same metrics on a number of publicly available neural audio codecs including Encodec~\cite{Encodec}, DAC~\cite{DAC} and AudioGen~\cite{audiogen}. Encodec and DAC are selected because are widely used for generative music modeling~\cite{copet2023simple, agostinelli2023musiclm, magnet}. We select the Encodec 32kHz variant because our MusicGen baseline relies on it, and DAC 44.1kHz because its alternatives operate at 24kHz and 16kHz. 
Further, since our autoencoder relies on a continuous latent, we also compare against AudioGen, a state-of-the-art autoencoder with a continuous latent. 
Notably, AudioGen presents both continuous and discrete options, and we report both for completeness.
All neural audio codecs are stereo, except DAC 44.1kHz and Encodec 32 kHz. In those cases, we independently project left and right channels and reconstruct from those.

The results in Table~\ref{tab:autoencoder} show that the proposed autoencoder is comparable or marginally worse in raw reconstruction quality with respect to the other available baselines, whilst targeting a significantly larger amount (2x-5x) of temporal downsampling, and hence a lower latent rate. Our results are not strictly comparable against discrete neural audio codecs, but are included as a reference. For a qualitative assessment of our autoencoder's reconstruction quality, listen to some examples on our demo site.

\subsection{Musical structure analysis}\label{sec:structure}

\begin{figure}[h]
     \centering
    \begin{subfigure}[t]{0.48\textwidth}
        \raisebox{-\height}{\includegraphics[width=0.18\textwidth]{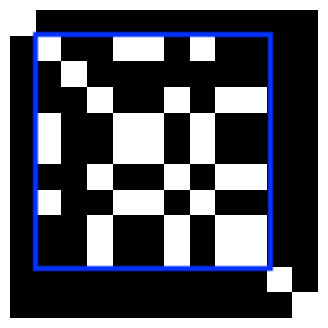}}
        \raisebox{-\height}{\includegraphics[width=0.18\textwidth]{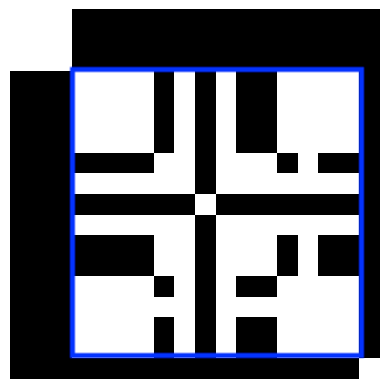}}
        \raisebox{-\height}{\includegraphics[width=0.18\textwidth]{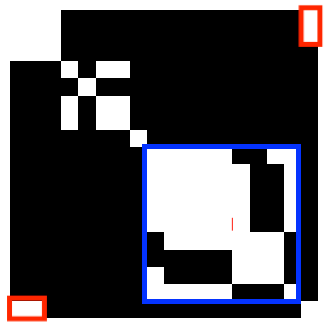}}
        \raisebox{-\height}{\includegraphics[width=0.18\textwidth]{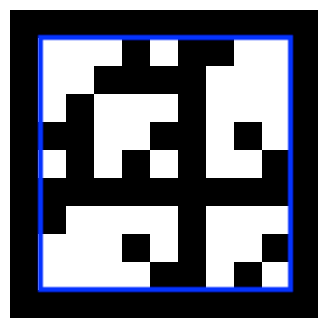}}
        \raisebox{-\height}{\includegraphics[width=0.18\textwidth]{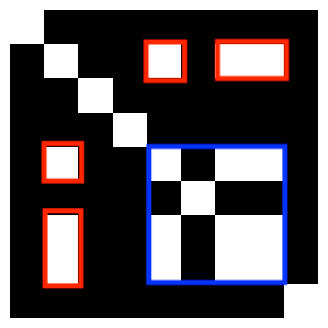}}
    \end{subfigure}    
    \begin{subfigure}[t]{0.48\textwidth}
        \raisebox{-\height}{\includegraphics[width=0.18\textwidth]{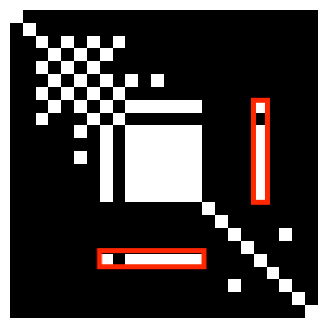}}
        \raisebox{-\height}{\includegraphics[width=0.18\textwidth]{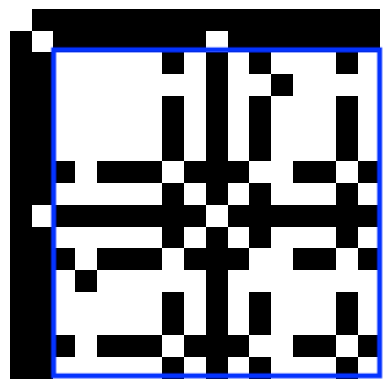}}
        \raisebox{-\height}{\includegraphics[width=0.18\textwidth]{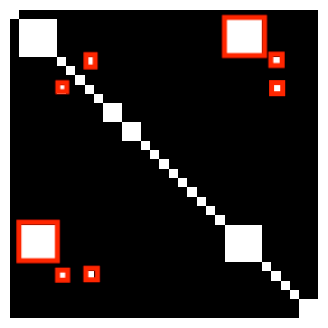}}
        \raisebox{-\height}{\includegraphics[width=0.18\textwidth]{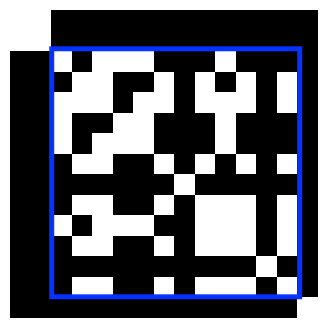}}
        \raisebox{-\height}{\includegraphics[width=0.18\textwidth]{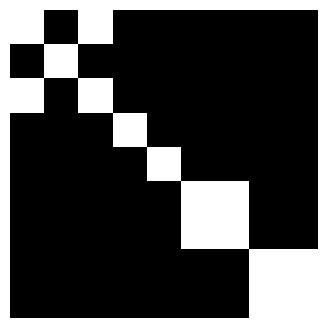}}
        \caption{SSMs of real music.}
        \label{fig:ssm_real}
        \vspace{1mm}
    \end{subfigure}    
    \begin{subfigure}[t]{0.48\textwidth}
        \raisebox{-\height}{\includegraphics[width=0.18\textwidth]{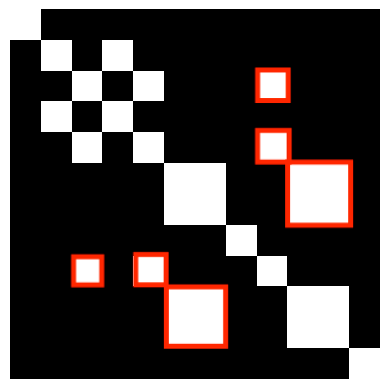}}
        \raisebox{-\height}{\includegraphics[width=0.18\textwidth]{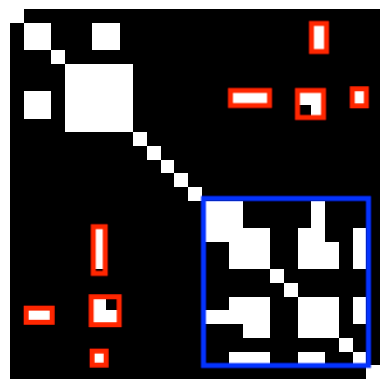}}
        \raisebox{-\height}{\includegraphics[width=0.18\textwidth]{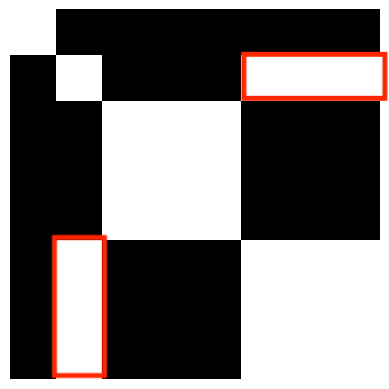}}       
        \raisebox{-\height}{\includegraphics[width=0.18\textwidth]{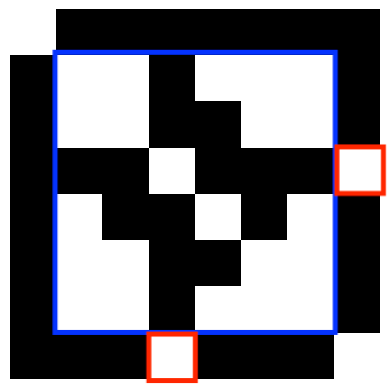}}
        \raisebox{-\height}{\includegraphics[width=0.18\textwidth]{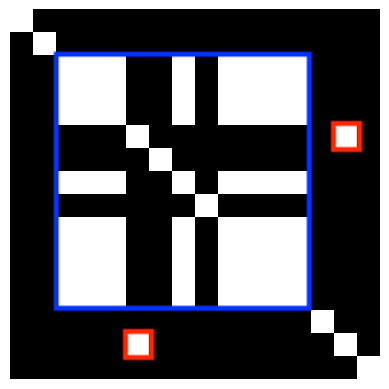}}
    \end{subfigure}    
    \begin{subfigure}[t]{0.48\textwidth}
        \raisebox{-\height}{\includegraphics[width=0.18\textwidth]{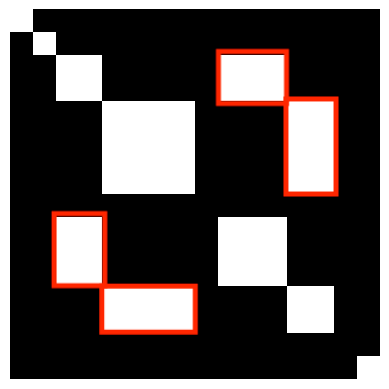}}
        \raisebox{-\height}{\includegraphics[width=0.18\textwidth]{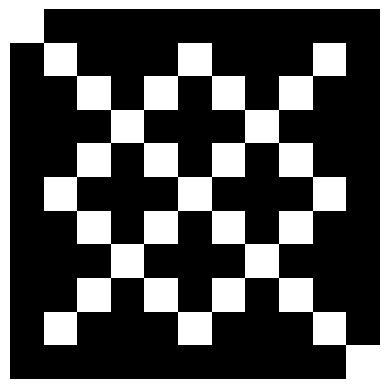}}
        \raisebox{-\height}{\includegraphics[width=0.18\textwidth]{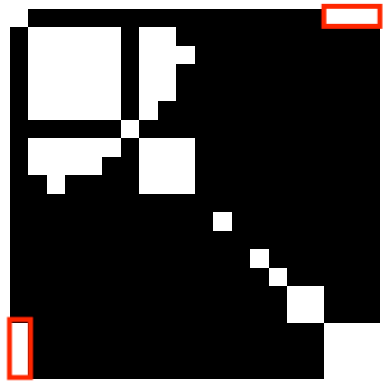}}       
        \raisebox{-\height}{\includegraphics[width=0.18\textwidth]{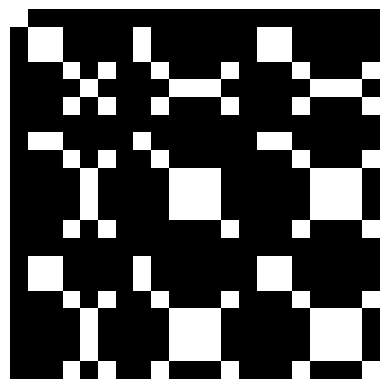}}
        \raisebox{-\height}{\includegraphics[width=0.18\textwidth]{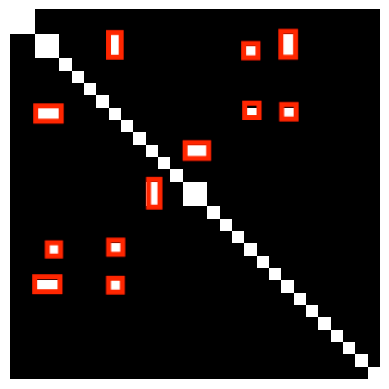}}
        \caption{SSMs of our model's generations.}
        \label{fig:ssm_stableaudio}
        \vspace{1mm}
    \end{subfigure}
    \begin{subfigure}[t]{0.48\textwidth}
        \raisebox{-\height}{\includegraphics[width=0.18\textwidth]{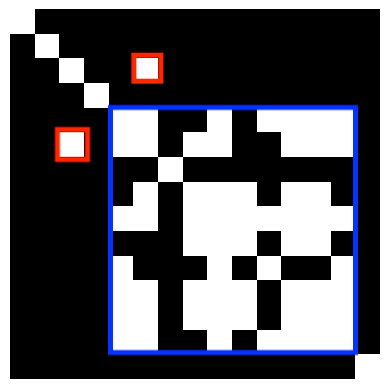}}
        \raisebox{-\height}{\includegraphics[width=0.18\textwidth]{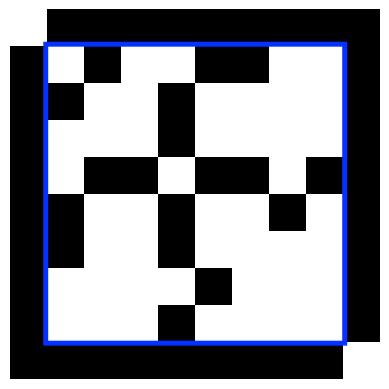}}
        \raisebox{-\height}{\includegraphics[width=0.18\textwidth]{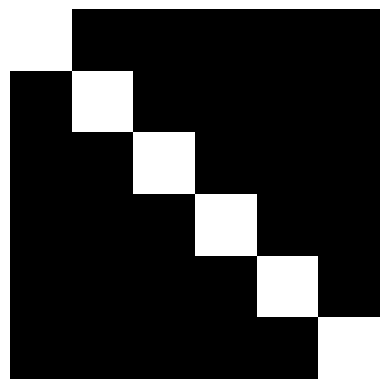}}
        \raisebox{-\height}{\includegraphics[width=0.18\textwidth]{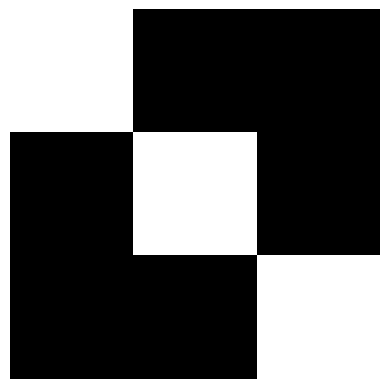}}
        \raisebox{-\height}{\includegraphics[width=0.18\textwidth]{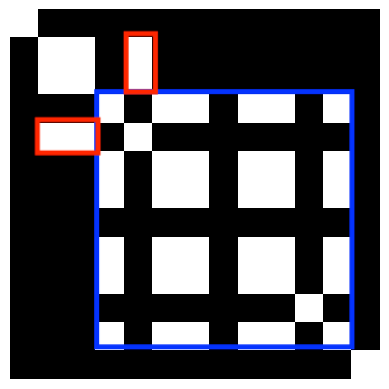}}
        \vspace{1mm}
        \caption{SSMs of MusicGen-large-stereo generations.}
        \label{fig:ssm_musicgen}
        \vspace{-2mm}
    \end{subfigure}    
    \caption{Each column shows the SSMs of different genres (left to right): rock, pop, jazz, hip-hop, and classical.}
    \label{fig:ssm}
    \vspace{-5mm}
\end{figure}

We explore the plausibility of the generated structures by visualizing the binary self-similarity matrices (SSMs) \cite{serra2014unsupervised} of randomly chosen generated music against real music of the same genre. Real music is from the Free Music Archive (FMA) \cite{defferrard2016fma}. 
Similarly to real music, our model's generations can build structure with intricate shifts, including repetition of motives that were introduced at the first section.
Red marks in Figure \ref{fig:ssm} show late sections that are similar to early sections.
In MusicGen examples, early sections rarely repeat (e.g., see diagonal lines in Figure \ref{fig:ssm_musicgen}) or music gets stuck in a middle/ending section loop (repetitive/loop sections are marked in blue in Figure \ref{fig:ssm}). Note that our model’s middle sections can also be repetitive, while still maintaining an intro/outro. 
We omit MusicGen's second row because most of its SSMs exhibit a similar behaviour.


\subsection{Memorization analysis}
Recent works \cite{carlini2023extracting, esser2024scaling} examined the potential of generative models to memorize training data, especially for repeated elements in the training set. Further, musicLM~\cite{agostinelli2023musiclm} conducted a memorization analysis to address concerns on the potential misappropriation of creative content. Adhering to principles of responsible model development, we also run a comprehensive study on memorization \cite{carlini2023extracting,agostinelli2023musiclm,esser2024scaling}. 

Considering the increased probability of memorizing repeated music within the dataset, we start by studying if our training set contains repeated data.
We embed all our training data using the LAION-CLAP$^8$ audio encoder to select audios that are close in this space based on a manually set threshold. The threshold is set such that the selected audios correspond to exact replicas. 
With this process, we identify 5566 repeated audios in our training set.

We compare our model's generations against the training set in LAION-CLAP$^8$ space. Generations are from 5566 prompts within the repeated training data (in-distribution), and 586 prompts from the Song Describer Dataset (no-singing, out-of-distribution). We then identify the top-50 generated music that is closest to the training data and listen. We extensively listened to potential memorization candidates, and could not find memorization. We even selected additional outstanding generations, and could not find memorization. The most interesting memorization candidates, together with their closest training data, are online for listening on our demo page.

\vspace{-1.5mm}
\subsection{Additional creative capabilities}
Besides text-conditioned long-form music generation, our model exhibits capabilities in other applications. While we do not conduct a thorough evaluation of these, we briefly describe those and showcase examples on our demo page.

\textit{Audio-to-audio} --- With diffusion models is possible to perform some degree of style-transfer by initializing the noise with audio during sampling~\cite{rouard2021crash,pascual2023full}. This capability can be used to modify the aesthetics of an existing recording based on a given text prompt, whilst maintaining the reference audio's structure (e.g.,~a beatbox recording could be style-transfered to produce realistic-sounding drums). As a result, our model can be influenced by not only text prompts but also audio inputs, enhancing its controllability and expressiveness. We noted that when initialized with voice recordings (such as beatbox or onomatopoeias), there is a sensation of control akin to an instrument. Examples of audio-to-audio are on our demo page.

\textit{Vocal music} --- The training dataset contains a subset of music with vocals. Our focus is on the generation of instrumental music, so we do not provide any conditioning based on lyrics. As a result, when the model is prompted for vocals, the model's generations contains vocal-like melodies without intelligible words. Whilst not a substitute for intelligible vocals, these sounds have an artistic and textural value of their own. Examples are given on our demo page.

\textit{Short-form audio generation} --- The training set does not exclusively contain long-form music. It also contains shorter sounds like sound effects or instrument samples. As a consequence, our model is also capable of producing such sounds when prompted appropriately. Examples of short-form audio generations are also on our demo page.


\vspace{-1.5mm}
\section{Conclusions}

We presented an approach to building a text-conditioned music generation model, operating at long enough context lengths to encompass full musical tracks. To achieve this we train an autoencoder which compresses significantly more in the temporal dimension than previous work. We model full musical tracks represented in the latent space of this autoencoder via a diffusion approach, utilizing a diffusion-transformer. We evaluate the trained model via qualitative and quantitative tests, and show that it is able to produce coherent music with state-of-the-art results over the target temporal context of 4m\,45s. 

\section{Ethics Statement}

Our technology represents an advancement towards aiding humans in music production tasks, facilitating the creation of variable-length, long-form stereo music based on textual input. This advancement greatly enhances the creative repertoire available to artists and content creators. However, despite its numerous advantages, it also brings inherent risks. A key concern lies in the potential reflection of biases inherent in the training data. Additionally, the nuanced context embedded within music emphasizes the necessity for careful consideration and collaboration with stakeholders. In light of these concerns, we are dedicated to ongoing research and collaboration with those stakeholders, including artists and data providers, to navigate this new terrain responsibly.
Adhering to best practices in responsible model development, we conducted an exhaustive study on memorization. Employing our methodology, we found no instances of memorization.


\bibliography{sa2}

%
%
%
%
%

\end{document}